\documentclass[twocolumn,floatfix,showpacs,prl,aps,amsmath]{revtex4}
\usepackage{bbold}
\usepackage{graphicx}

\newcommand{\be}{\begin{equation}}
\newcommand{\en}{\end{equation}}
\newcommand{\avg}[1]{\left< #1 \right>}

\newcommand{\diag}{{\rm diag}}
\newcommand{\tr}{{\rm Tr}}

\newcommand{\bn}{\boldsymbol{n}}
\newcommand{\bm}{\boldsymbol{m}}

\newcommand{\hq}{\hat{q}}

\newcommand{\phin}{\Phi_{n}}
\newcommand{\thetan}{\Theta_{n+1}}

\newcommand{\tras}{\avg{\tr A^2}}

\newcommand{\dz}{\delta Z}

\newcommand{\ttau}{\tilde{\tau}}
\newcommand{\ptt}{P(\{\ttau_i\})}
\newcommand{\ptn}{P_n(\{\ttau_i\})}
\newcommand{\ptnpo}{P_{n+1}(\{\ttau_i\})}
\newcommand{\dtau}{\delta \ttau}
\newcommand{\ttaun}{\ttau^{(n)}}
\newcommand{\ttaunpo}{\ttau^{(n+1)}}

\newcommand{\brai}{\langle i|}
\newcommand{\brak}{\langle k|}
\newcommand{\keti}{|i \rangle}

\newcommand{\rst}{\rho_{st}}

\newcommand{\dell}{\delta L}

\newcommand{\vmat}{\overline{V}}
\newcommand{\br}{\boldsymbol{r}}
\newcommand{\brho}{\boldsymbol{\rho}}

\newcommand{\prodrho}{\prod_{i=1}^{d-1}}

\newcommand{\lfermi}{\lambda_F}

\begin{document}
\title{Scattering approach to Anderson localisation}
\author{A. Ossipov}
\email{alexander.ossipov@nottingham.ac.uk}
\affiliation{School of Mathematical Sciences, University of Nottingham, Nottingham NG7 2RD, United Kingdom} 
\date{\today}
\begin{abstract}
We develop a novel approach to the Anderson localisation problem in a $d$-dimensional disordered sample of dimension  $L\times M^{d-1}$. Attaching  a perfect lead with the cross-section $M^{d-1}$ to one side of the sample, we derive evolution equations for the scattering matrix and the Wigner-Smith time delay matrix as a function of $L$.  Using them one obtains the Fokker-Planck equation for the distribution of the proper delay times and the evolution equation for their density at weak disorder. The latter can be mapped onto a non-linear partial differential equation of the Burgers type,  for which a complete  analytical solution for arbitrary $L$ is constructed. Analysing the solution for a cubic sample with $M=L$ in the limit $L\to \infty$, we find that for $d<2$ the solution tends to the localised fixed point, while for $d>2$ to the metallic fixed point and provide explicit results for the density of the delay times in these two limits. 
\end{abstract}


\maketitle

\emph{Introduction--}
Sixty years ago Anderson discovered that the classical diffusion in a random potential can be totally suppressed by quantum interference effects \cite{And58}. Since that time the problem of Anderson localisation has remained in the focus of very active research and recently it has received a lot of attention in the context of topological insulators and many-body localisation \cite{Abrahams}. 

Apart from the strictly one-dimensional case, the most developed non-perturbative theory of Anderson localisation is available for disordered wires. The only important parameters of such a system are the length $L$, the mean free path $l$, the number of the propagating modes $N$ at the Fermi energy $E$ and the localisation length $\xi=Nl$. The disorder is usually assumed to be weak, so that $L,l\gg \lambda_F$, where $\lambda_F$ is the Fermi wave length.
There are two powerful analytical approaches, which can solve the problem of Anderson localisation in a wire for an arbitrary ratio  $L/\xi$: the Dorokhov, Mello, Pereyra and Kumar  (DMPK) equation \cite{D82, MPK88} and the supersymmetric nonlinear $\sigma$-model \cite{EL83, FM92}. Both solutions are restricted to the quasi-one-dimensional geometry of a wire, for which the transverse dimension $M$ is much smaller than $L$. Despite a lot of efforts, a similar theory for higher dimensional systems has not been developed so far and it is the purpose of this Letter to take the first step towards this long-standing goal.

We consider a $d$-dimensional weakly disordered sample of the length $L$ in the $x$-direction and the width $M$ in all other transverse directions. A perfect lead is attached to one side of the sample along the $x$-direction, which has the same cross section $M^{d-1}$ as the sample. The scattering setup allows one to introduce the scattering matrix $S$ and the Wigner-Smith time-delay matrix $Q=-i\hbar S^{-1/2}\frac{\partial S}{\partial E}S^{-1/2}$, whose eigenvalues $\ttau_i$ are referred to as the proper delay times (see Ref.\cite{T16,K05} for reviews). Generalising the approach developed for the one-dimensional systems \cite{OKG00, ACT97} we derive the Fokker-Planck equation for the evolution of the distribution function  $P(\{\ttau_i\},r)$ in fictitious time $r\propto L/l$, provided that $L,M, l\gg \lfermi$. Then we focus on the time-dependent equation for the density $\rho(\ttau,r)$ of the delay times, which contains important information about localisation in the corresponding closed system. Mapping this equation onto a non-linear partial differential equation of the Burgers type, we construct its complete analytical solution for  arbitrary $L$, $M$ and $l$.   

Our general solution, which is valid for any dimensionality $d$, allows us in particular to consider a $d$-dimensional cubic sample with $M=L$. Analysing such a system in the limit $L/\lambda_F\to \infty$, we find that for $d<2$ the solution tends to the localised fixed point, while for $d>2$ to the metallic (diffusive) fixed point and derive explicit analytical results for the density of the delay times in these two limits. Thus our approach provides a solid non-perturbative foundation for the arguments of the scaling theory of Anderson localisation \cite{AALR79}. 

As the derivation of our results involves a lot of technical steps, in this Letter we only outline its main points 
and leave the technical details for a more specialised publication \cite{Ossipov}.


\emph{Model--}
We consider the Hamiltonian for a particle moving in the $d$-dimensional $\delta$-correlated disordered potential:
\begin{eqnarray}
&&H=-\sum_{i=0}^{d-1}\frac{\partial^2}{\partial x_i^2}+V(\br),\; \br=(x,\brho),\nonumber\\
&&\avg{V(\br)V(\br')}=\sigma \delta(\br-\br'),\;  \sigma=\frac{1}{2\pi \nu \tau_s},
\end{eqnarray}
where $x\equiv x_0$, $\brho\equiv (x_1,\dots x_{d-1})$, $\nu$ is the density of states, $\tau_s$ is the scattering mean free time and we set $\hbar=2m=1$. A sample is assumed to be finite with $-L\le x \le 0$ and $0\le x_i \le M$ for $i=1,\dots, d-1$, and the Dirichlet boundary condition is imposed in all directions. 

By attaching a perfect lead to one side of the sample at $x=0$, we obtain a scattering system characterised by the $N\times N$ $S$-matrix, which is unitary $S^{\dagger}=S^{-1}$ and symmetric $S^{T}=S$ due to the time reversal symmetry. The eigenfunctions in the transverse directions
$u_{\bn}(\brho)=\left(\frac{2}{M}\right)^{\frac{d-1}{2}}\prodrho\sin \frac{\pi n_i x_i}{M},\, n_i\in \mathbb{N}$, 
correspond to the eigenenergies $E_{\bn}=\left(\frac{\pi \bn }{M}\right)^2$. The number of open channels at the energy $E$ is equal to $N=\gamma_{d-1}(M\sqrt{E}/\pi)^{d-1}$, where 
$\gamma_d=\frac{\pi ^{\frac{d}{2}}}{2^d\Gamma\left(\frac{d}{2}+1\right)}$.  


\emph{Recursion relations for  $S$ and $Q$ matrices--}
In order to derive an equation for the evolution of  $S$ by increasing $L$ to $L+\dell$, we first consider the scattering matrix of a thin slice of the length $\dell \ll \lfermi$. Using the Lippmann-Schwinger equation, one can show that the reflection and the transmission matrices from the left and from the right coincide respectively, $r'=r$, $t'=t$, and to the leading order in $\dell/\lfermi$ are given by
\begin{eqnarray}
r=-B(I+B)^{-1}, \; t=I+r,\; B\equiv \frac{i}{2}\hq^{-1/2}\vmat(0) \hq^{-1/2},
\end{eqnarray}
where $\hq$ is the diagonal matrix, whose elements are the quantised longitudinal momenta $q_{\bn}=\sqrt{E-E_{\bn}}$ and $\overline{V}_{\bn\bm}(x)\equiv \int_x^{x+\dell}dx' \int d\brho V(x',\brho) u_{\bn}(\brho)u_{\bm}(\brho)$.

Applying the standard formula for the composition of the scattering matrices and using the fact that $r=t-I$ we derive the relation between $S_{n+1}\equiv S(L+\dell)$ and $S_n\equiv S(L)$, the scattering matrices corresponding to the system of the length $L=n\dell $ and $L+\dell=(n+1)\dell$:
\begin{eqnarray}\label{S-recursion}
f\left(S_{n+1}\right)=f\left(e^{i\hq \dell}S_{n}e^{i\hq \dell}\right)+A_{n+1},
\end{eqnarray}
where $A_{n+1}\equiv \hq^{-1/2}\vmat(L) \hq^{-1/2},$ and $f(S)\equiv i\frac{(S-I)}{(S+I)}$. The above equation is a direct generalisation of the one-dimensional relation \cite{OKG00}. Differentiating it w.r.t. $E$ one obtains the recursion relation for $Q$:
\begin{eqnarray}\label{Q-recursion}
W_{n}Q_{n+1}W_{n}^T&=&
C_n
\left(J_nQ_nJ_n^T+K_n\right)C_n+H_n.
\end{eqnarray}
All the matrices involved in this equation can be expressed through $S_n$, $A_{n+1}$ and $\hq$ and their definitions are given in the Supplemental Material \cite{suppl}. Both relations preserve the symmetries of the scattering and Wigner-Smith matrices respectively: $S^{\dagger}=S^{-1}$, $S^{T}=S$, $Q^{\dagger}=Q$, $Q^{T}=Q$. They hold for any strength of disorder $\sigma$ and are very convenient for numerical simulations, as they deal with the matrices corresponding to  $d-1$  rather than $d$-dimensional systems. 

Now we assume that disorder is weak, i.e. $l\gg \lfermi$. Then an analysis of the relations \eqref{S-recursion} and \eqref{Q-recursion} suggests that the change of $S$ and the eigenvectors of $Q$ at each step of the recursion is governed by the parameter $\dell/\lambda_F$, while the change of the eigenvalues of $Q$ by the parameter $\dell/l$ \cite{suppl}. As $\dell/\lambda_F \gg \dell/l$, this implies that $S$ and the matrix of the eigenvectors of $Q$, $O$, represent fast variables, while $\ttau_i$ are  slow variables. Therefore in the following we assume that for $L\gg\lfermi$, $S$ and $O$ are statistically independent random matrices and the first two moments of the distribution of their matrix elements satisfy the following conditions
\begin{eqnarray}
&&\avg{S_{ij}}=0,\quad \avg{O_{ij}}=0,\quad  \avg{S_{ij}S_{kl}}=0, \nonumber\\  
&& \avg{S_{ij}S_{kl}^*}=\frac{\delta_{ik}\delta_{jl}+\delta_{il}\delta_{jk}}{N+1},
\;\avg{O_{ij}O_{kl}}=\frac{\delta_{ik}\delta_{jl}}{N}.
\end{eqnarray}
These relations can be justified by two observations: (i) the phases of the $S$-matrix elements are fast oscillating even in the absence of disorder, (ii) the momentum of a reflected particle is completely randomised for weak disorder. In Supplemental Material we explain why these conditions are strongly  motivated by the recursion relations and check their validity by numerical simulations.


\emph{Fokker-Planck equation and the evolution equation for the density --}
The recursion relation \eqref{Q-recursion} can be transformed into the Fokker-Planck equation for the joint probability distribution function $\ptt$ in the continuum limit $\dell\to 0$. To this end, we first use the general relation between $\ptt$ calculated at two consequent steps:
\begin{eqnarray}\label{pln-init}
&&\ptnpo=\ptn+ \\
&&\left[
-\sum_i \frac{\partial}{\partial \ttau_i}\avg{\dtau_i}
+\frac{1}{2}\sum_{ik}\frac{\partial^2}{\partial \ttau_i \partial\ttau_k}\avg{\dtau_i\dtau_k}
\right]\ptn,\nonumber
\end{eqnarray}
where $\avg{\dots}$ stands for the averaging over $S$, $O$ and $V(\br)$ and only the terms up to the first order in $\dell$ must be retained on the r.h.s..The averages $\avg{\dtau_i}$ and $\avg{\dtau_i\dtau_k}$ can be computed with the help of the perturbation theory:
\begin{eqnarray}
\dtau_i=\brai O^T\delta Q O\keti +\sum_{k\neq i}\frac{|\brak O^T\delta Q O\keti |^2}{\ttau_i-\ttau_k},
\end{eqnarray}
where $\{\keti\}$ is the standard basis in $\mathbb{R}^N$ and we omit the index $n$ for all variables to lighten the notation. The matrix $\delta Q\equiv Q_{n+1}-Q_n$ can be found from Eq.\eqref{Q-recursion}.

Introducing the scaled variables $\tau=\frac{\ttau}{\tau_s}$ and  $r=A_d\frac{L}{l}$, with $A_d\equiv \frac{\sqrt{\pi}\Gamma\left(\frac{d+1}{2}\right)}{ \Gamma\left(\frac{d}{2}\right)}$, and taking the limit $\dell\to 0$,
we derive (see the Supplemental Material \cite{suppl} for details) the Fokker-Planck equation for the distribution function
$P(\{\tau_i\},r)$
\begin{eqnarray}\label{FP-tau-final-const}
&&\frac{\partial P}{\partial r} =
\frac{1}{N}\sum_i \frac{\partial}{\partial \tau_i}
\left[(N-1)\tau_i-2N\right.\nonumber\\
&&\left.
-\sum_{k\neq i}\frac{\tau_i^2}{\tau_i-\tau_k}+\frac{\partial }{\partial \tau_i}
\tau_i^2\right]P.
\end{eqnarray}
The distribution function $P(\{\tau_i\},r)$ contains the full information about the delay times, however in order to distinguish between the localised and delocalised phases of the closed system, it is sufficient to study a simpler quantity -- the density of the delay times $\rho(\tau,r)=\frac{1}{N}\sum_i\avg{\delta(\tau-\tau_i)}$, which can be obtained from $P(\{\tau_i\},r)$ by integrating out all but one variables $\tau_i$. 

The evolution equation for $\rho(\tau,r)$, which can be derived from \eqref{FP-tau-final-const} in the standard way \cite{B97}, reads
\begin{eqnarray}\label{density-eq-scaled-3}
\frac{\partial \rho}{\partial r}=
\frac{\partial}{\partial \tau}\left[ \rho
\left(\tau-2-\tau^2\int d\tau' \frac{\rho(\tau',t)}{\tau-\tau'}\right)
+\frac{\partial}{\partial \tau}\frac{\tau^2\rho}{2N}\right].
\end{eqnarray}


\emph{Burgers equation and the stationary solution --}
The integro-differential equation for the density can be mapped onto a non-linear partial differential equation employing the method used in Ref.\cite{BRM94}. We introduce the Stieltjes transform of $\rho(\tau,r)$ defined as
\be
F(z,r)=\int_0^{\infty}d\tau' \frac{\rho(\tau',r)}{z-\tau'}.
\en
The function $F(z,r)$ is analytic in the complex plane for all $z$ except the positive real axis, where it is discontinuous:
\be
F_{\pm}\equiv \lim_{\epsilon\to 0^+}F(\tau\pm i\epsilon)=
\pm \frac{\pi}{i}\rho (\tau,r)+\int_0^{\infty}d\tau' \frac{\rho(\tau',r)}{\tau-\tau'}.
\en
Using this formula, the analyticity of $F$ and Eq.\eqref{density-eq-scaled-3} one can show that $F$ satisfies the non-linear differential equation
of the Burgers type
\be\label{F-eq-full-alt}
\frac{\partial F}{\partial r}=
\frac{1}{2N}\frac{\partial}{\partial z}
\left[N \left( 2\left(z-2\right)F -z^2F^2\right)
+\frac{\partial }{\partial z}z^2F\right],
\en
whose solution allows us to find $\rho$ through the relation $\rho(\tau,r)=\frac{i}{2\pi}(F_+-F_-)$. 


\emph{Hopf-Cole transformation and the non-stationary solution --}
In order to find a solution of Eq.\eqref{F-eq-full-alt} we employ a variant of the Hopf-Cole transformation:
\begin{eqnarray}\label{Hopf-Cole-y}
\label{F-u-relation}F(z,r)&=&\frac{z-2}{z^2}-\frac{4}{z^2}\frac{u'_s(s,r)}{u(s,r)},\; s=-\frac{4N}{z}
\end{eqnarray}
which maps the equation for $F$ onto the generalised diffusion equation:
\begin{eqnarray}\label{diffusion-eq-u}
8N \frac{\partial u}{\partial r}=4s^2u''_{ss}-s(s+4N)u.
\end{eqnarray}

One can look for the general solution of this equation as a linear combination of the eigenfunctions $e^{-\frac{\lambda}{2} r}u_{\lambda}(s)$.
It turns out that the spectrum is continuous for $\lambda =\frac{4\mu^2+1}{4N}$, and the corresponding orthogonal eigenfunctions are given by the Whittaker functions  $W_{-N,i\mu}(s)$ with $\mu>0$ \cite{SB10}. Additionally to this set of the eigenstates there is another eigenfunction $W_{-N,\frac{1}{2}}(s)$ for $\lambda=0$ corresponding to the stationary state \cite{comment}. Thus the solution of Eq.\eqref{diffusion-eq-u} can be written as
\begin{eqnarray}\label{u_sol_prelim}
u(s,r)=c_0W_{-N,\frac{1}{2}}(s)+\int_0^{\infty}d\mu\, c(\mu)e^{-\frac{(4\mu^2+1)r}{8N}}W_{-N,i\mu}(s),\nonumber\\
\end{eqnarray}
where the coefficients $c_0=\Gamma(N+1)$ and $c(\mu)=\frac{8\mu\sinh(\pi \mu)\Gamma\left(N+\frac{1}{2}+i\mu \right)\Gamma\left(N+\frac{1}{2}-i\mu \right)}{\pi (1+4\mu^2)\Gamma (N)}$ are determined from the initial condition $u(s,0)=e^{-\frac{s}{2}}$. This formula along with Eq.\eqref{F-u-relation} and the relation $\rho(\tau,r)=\frac{i}{2\pi}(F_+-F_-)$ provides the general solution for $\rho(\tau,r)$, which is valid for any $L/\lfermi\gg 1$, $N\propto (M/\lfermi)^{d-1} \gg 1$ and $l/\lfermi \gg 1$.


\emph{The density of delay times for a cubic sample in the thermodynamic limit --}
For a cubic sample $M=L$ and it follows from Eq.\eqref{u_sol_prelim} that the $r$-dependence of the solution is governed by the parameter $r/N\propto (\lfermi/l)(L/\lfermi)^{2-d}$, which has a meaning of the inverse dimensionless conductance $g^{-1}$. One can see that in the thermodynamic limit ($L/\lfermi\to \infty$), $r/N \to \infty$ for $d<2$  and $r/N \to 0$ for $d>2$. In the former case, the solution tends to its localised fix point given by $W_{-N,\frac{1}{2}}(s)$, whereas in the latter case it tends to the metallic (diffusive) fixed point, where the contribution from all $W_{-N,i\mu}(s)$ is important. The $d=2$ case is a marginal one and requires more careful treatment \cite{Ossipov}.


\emph{Localised regime --}
In the localised regime, where the solution is determined by the stationary state, the density can be found from the asymptotics of $W_{-N,\frac{1}{2}}(s)$ at $N\to \infty$. As $s \propto N/\tau$, such asymptotics depend generally on the value of $\tau$. It turns out, that one needs to consider separately two different regimes: $\tau\sim N^{0}$ and  $\tau\sim N^2$, for which the asymptotics of $W_{-N,\frac{1}{2}}(s)$ and hence the expressions for the density are different:
\begin{eqnarray}\label{density-localised}
\rst(\tau)=\begin{cases} \frac{2}{\pi}\frac{\sqrt{\tau-1}}{\tau^2} &, \quad \tau\sim N^0,\; \tau\ge 1\\
\frac{4N}{\tau^2} &, \quad  \tau\gtrsim N^{2}.
\end{cases}
\end{eqnarray}
A long  $\tau^{-2}$ tail in the distribution of the delay times in the localised regime was previously found analytically for $1d$ and quasi-$1d$ systems \cite{T16,F03}. In the numerical simulations for the $2d$ Anderson model both power-laws $\tau^{-\frac{3}{2}}$ and $\tau^{-2}$, which follow from our result, were identified \cite{XW11}.

The localisation length can be estimated as $\xi \propto v_F \tau_W^{typ}$, where $v_F$ is the Fermi velocity and $\tau_W^{typ}$ is a typical value of the Wigner delay time $\tau_W=\sum_{i=1}^N\ttau_i$. According to Eq.\eqref{density-localised} a typical value of $\ttau$ is of order of $\tau_s$ and therefore $\xi \propto N v_F \tau_s=Nl$. This result is in agreement with the quasi-1d result, where $L\to \infty$ at constant $W$. For a cubic sample with $d<2$, $N\propto (L\sqrt{E})^{d-1} $ grows with $L$, however $\xi/L\to 0$ in the thermodynamic limit, as expected in the localised regime.


\emph{Diffusive and ballistic regimes --} In the metallic regime, where $r/N\ll 1$, a direct analysis of Eq.\eqref{u_sol_prelim} is complicated, so it is more convenient to derive the limiting solution in a different way. For $r/N\ll 1$ the last term in Eq.\eqref{F-eq-full-alt} is small and hence can be neglected, then introducing the new function $\psi(\xi,r)$, such that $F=\frac{z-2}{z^2}+z^{-1}\psi(\ln z,r)$, one can map Eq.\eqref{F-eq-full-alt} onto the inviscid forced Burgers equation
\begin{eqnarray}
\frac{\partial \psi}{\partial r}+\psi\frac{\partial \psi}{\partial \xi} =2e^{-\xi}-4e^{-2\xi},
\end{eqnarray}
which can be solved by the method of characteristics:
\begin{eqnarray}\label{F-diffusive}
F(z,r)=
\frac{z-2+ 2\sqrt{1-z+\frac{z^2}{z_0^2}}}{z^2},
\end{eqnarray}
where $z_0=z_0(z,r)$ is determined implicitly by the equation $f(z_0,r)=z$ with $f(x,r)\equiv \frac{x}{2}\left(x\left(1+\cosh\frac{2r}{x}\right)+2\sinh\frac{2r}{x}\right)$. This formula gives a solution at an arbitrary value of $r\propto L/l$ in the metallic regime. Now we can analyse it in detail in the ballistic ($L/l\ll 1$) and the diffusive  ($L/l\gg 1$) limits.

In the ballistic regime, $r\ll 1$, one can expand $f(x,r)$ in the power-series in $r/x$ and find $z_0$ approximately. The leading order result reads:
\begin{eqnarray}
F(z,r)\approx \frac{1}{z-2r},\; \Rightarrow \rho(\tau,r)=\delta (\tau-2r),
\end{eqnarray}
which describes a ballistic motion with the Fermi velocity, $L\propto v_{F} \ttau$, as expected.

In the diffusive regime ($r\gg 1$), the solution can be found by scaling $z_0=yr$, $z=wr^2$ and  $F(z,r)= \frac{1}{r^2}\tilde{F}(\frac{z}{r^2},r)$ and keeping only the leading order terms in $r$. The appearance of such a scaling implies that a typical delay time $\ttau\propto L^2/D$ ($D$ is the classical diffusion constant), which is very natural in the diffusive regime. The function $\tilde{F}(w,r)$ is then given by
\begin{eqnarray}
\label{tilde-F}\tilde{F}(w,r)=
 \frac{1}{w}+\frac{2}{rw^{\frac{3}{2}}}\sqrt{\frac{w}{y^2}-1},
\end{eqnarray}
where $y=y(w)$ satisfies the equation $y\cosh y^{-1}=\sqrt{w}$. This result implies that $\rho(\tau,r)\approx \tilde{\rho}(w)/r^3\neq 0$ only for $w\in [w_{min},w_{max}]$, where $w_{min}\approx \frac{\pi^2}{16r^2}$ and $w_{max}\approx 2.28$. The behaviour of $\tilde{\rho}(w)$ can be found analytically at $w\to w_{min}$, where $\tilde{\rho}(w)\approx \frac{2}{\pi w^{\frac{3}{2}}}$, and at $w\to w_{max}$, where $\tilde{\rho}(w)\approx \frac{2\sqrt{(w_{max}+1)(w_{max}-w)}}{\pi w_{max}^2}$. For intermediate values of $w$,   $\tilde{\rho}(w)$ can be determined numerically from Eq.\eqref{tilde-F}. 

The appearance of the power-law $\tau^{-\frac{3}{2}}$ tail in the metallic regime can be related to the classical diffusion \cite{OKG03}.


\emph{Comparison with the DMPK equation and other approaches--}
Since our method works also for a quasi-$1d$ geometry, it makes sense to compare it with the DMPK equation. In Refs. \cite{BPB96, BC96} the DMPK equation for the reflection eigenvalues in the presence of absorption was derived. As the proper delay times can be extracted from the reflection eigenvalues in the limit of weak absorption \cite{B01}, one can obtain the DMPK equation for proper delay times and compare it with our Eq.\eqref{FP-tau-final-const}. It turns out that  Eq.\eqref{FP-tau-final-const} coincides with the DMPK equation in the quasi-$1d$  case.

We stress that the scattering isotropy assumption for a thin slice, which is crucial for the derivation of the DMPK equation \cite{B97}, is not used in our approach, in which the scattering properties of a slice are treated microscopically. This allows us to study the problem in higher dimensions.

In Ref.\cite{OF05} a similar scattering setup with a single multi-channel lead was considered and a relation between the statistics of the partial delay times and certain correlation functions of the non-linear $\sigma$-model was derived. In contrast to the present method,  such an approach is limited by the available solutions of the $\sigma$-model: one can either employ a non-perturbative solution for the quasi-1d geometry or rely on the perturbative expansion in the metallic regime in higher dimensions. These limitations are shared by most of the other known methods, in contrast to our approach. 

Another outcome of Ref.\cite{OF05} is a simple relation between the statistics of the delay times and the local statistics of the wave functions derived for a single-channel lead. It would be of great interest to generalise that relation to a multi-channel case, this would allow one to get information about wave functions of a closed sample directly from the results of the present work.


\emph{Conclusions--} We have developed a new  approach to the $d$-dimensional Anderson localisation problem, which enabled us to obtain in a non-perturbative way the statistics of the delay times in the ballistic, diffusive and localised regimes at weak disorder. It overcomes the limitations of the existing methods and paves the way for studying analytically Anderson localisation in higher dimensional systems.

\acknowledgments

I acknowledge useful discussions with  C.~W.~J. Beenakker, P.~W. Brouwer, V. Cheianov, Y.~V. Fyodorov and C. Texier.

\begin{widetext}
\clearpage
\begin{center}
\textbf{\large Supplemental Material: Scattering approach to Anderson localisation}
\end{center}
\end{widetext}
\setcounter{equation}{0}
\setcounter{figure}{0}
\setcounter{table}{0}
\setcounter{section}{0}
\setcounter{page}{1}
\makeatletter
\renewcommand{\theequation}{S\arabic{equation}}

\section{Recursion relation for $Q$ matrices}

The recursion relation for the $S$-matrix reads
\begin{eqnarray}\label{S-recursion-suppl}
f\left(S_{n+1}\right)=f\left(e^{i\hq \dell}S_{n}e^{i\hq \dell}\right)+A_{n+1},
\end{eqnarray}
where $\hq$ is the diagonal matrix, whose elements are the longitudinal momenta $q_{\bn}=\sqrt{E-E_{\bn}}$,
$A_{n+1}\equiv \hq^{-1/2}\vmat(L) \hq^{-1/2}$ and $f(S)\equiv i\frac{(S-I)}{(S+I)}$. In order to derive the recursion relation for $Q_n$, it is convenient first to introduce the following notation
\begin{eqnarray}\label{T-def}
S_n\equiv e^{2i \Theta_n},\; T_n\equiv e^{i\hq \dell} S_{n}e^{i\hq \dell}\equiv e^{2i \Phi_n}.
\end{eqnarray}
Then the recursion relation can be written as
\begin{eqnarray}\label{tanphi}
\tan \thetan=\tan\phin-A_{n+1}.
\end{eqnarray}
Using that $f(S)=i I-2i (S+I)^{-1}$ and  differentiating Eq.\eqref{S-recursion-suppl} w.r.t. $E$ one obtains
\begin{eqnarray}\label{T-derivative}
&&S_{n+1}^{-1/2}\frac{dS_{n+1}}{dE}S_{n+1}^{-1/2}=
\Gamma_n^{-1} T_{n}^{-1/2}\frac{dT_{n}}{dE}T_{n}^{-1/2}\left(\Gamma_n^T\right)^{-1}\nonumber\\
&&-2i\cos\thetan\frac{d A_{n+1}}{dE}\cos\thetan,
\end{eqnarray}
where $\Gamma_{n}=\cos\phin\left(\cos\thetan\right)^{-1}$.
From Eq.\eqref{T-def} we find that
\begin{eqnarray}
\frac{dT_{n}}{dE}=e^{i\hq \dell}\left(i\frac{d \hq}{dE}S_{n}\dell+iS_{n}\frac{d \hq}{dE}\dell+
\frac{d S_{n}}{dE}\right)e^{i\hq \dell}.
\end{eqnarray}
Using this result and the fact that $\frac{d \hq}{dE}=(2 \hq)^{-1}$ we can rewrite the term $T_{n}^{-1/2}\frac{dT_{n}}{dE}
T_{n}^{-1/2}$ in Eq.\eqref{T-derivative} as
\begin{eqnarray}
&&\frac{i}{2}T_{n}^{-1/2}e^{i\hq \dell}\left( \hq^{-1}S_{n}+S_{n} \hq^{-1}\right)e^{i\hq \dell}T_{n}^{-1/2}\dell\nonumber\\
&&+iT_{n}^{-1/2}e^{i\hq \dell}S_{n}^{1/2}Q_{n}S_{n}^{1/2}e^{i\hq \dell}T_{n}^{-1/2},
\end{eqnarray}
where we introduced the Wigner-Smith time delay matrix $Q$ defined as
\begin{eqnarray}
Q=-i S^{-1/2}\frac{dS}{dE}S^{-1/2}.
\end{eqnarray}
Now we can rewrite Eq.\eqref{T-derivative} as a recursion relation for the $Q$-matrices:
\begin{eqnarray}
\label{Q-transform}Q_{n+1}&=&
\Gamma_n^{-1}\left(J_n Q_nJ_n^T+K_n\right)\left(\Gamma_n^T\right)^{-1}+R_n,
\end{eqnarray}
where the following matrices were introduced
\begin{eqnarray}
J_{n}&=&T_{n}^{-1/2}e^{i\hq \dell}S_{n}^{1/2}, \; J_{n}J_{n}^T=I,\; J_n^*=J_n,\nonumber\\
K_n&=&\frac{1}{2}  T_{n}^{-1/2}e^{i\hq \dell}\left( \hq^{-1}S_{n}+S_{n} \hq^{-1}\right)e^{i\hq \dell}T_{n}^{-1/2}\dell,\nonumber\\
 K_n^*&=&K_n, \; K_n^T=K_n \nonumber\\
R_n&=&-2\cos\thetan\frac{d A_{n+1}}{dE}\cos\thetan,\nonumber\\
R_n^*&=&R_n,\;R_n^T=R_n.\nonumber
\end{eqnarray}
It follows from the two symmetries of the scattering matrix $S^T=S$ and $S^{\dagger}=S^{-1}$, that $\Phi=\Phi^{\dagger}=\Phi^T$ and $\Theta=\Theta^{\dagger}=\Theta^T$, which implies that $\Gamma^T=\Gamma^{\dagger}$. Thus the transformation \eqref{Q-transform} preserves both symmetries of the $Q$-matrix $Q=Q^T$ and $Q=Q^{\dagger}$, as expected.

From Eq.\eqref{tanphi} one can find a useful representation for $\Gamma_n$.
\begin{eqnarray}\label{Delta-def}
\Gamma_n&=&(I+\Delta_n)^{1/2}W_n,\\
\Delta_n&\equiv& -\sin\phin A_{n+1}\cos\phin-\cos\phin A_{n+1}\sin\phin\nonumber\\
&+&\cos\phin A_{n+1}^2\cos\phin,\;\Delta_n^T=\Delta_n,\; \Delta_n^*=\Delta_n\nonumber\\
W_n&\equiv&(I+\Delta_n)^{-1/2} \cos\phin \nonumber\\
&\times&\left(\left(\cos\phin\right)^{-1}(I+\Delta_n)\left(\cos\phin\right)^{-1}\right)^{1/2},\nonumber\\
&&W_nW_n^T=I,\; W_n^*=W_n. \nonumber
\end{eqnarray}
One can use this expression for $\Gamma_n$ in Eq.\eqref{Q-transform} in order to rewrite it as
\begin{eqnarray}\label{Q-recursion-suppl}
W_{n}Q_{n+1}W_{n}^T=
C_n
\left(J_nQ_nJ_n^T+K_n\right)C_n+H_n,
\end{eqnarray}
where $C_n\equiv (I+\Delta_n)^{-1/2}$ and $H_n\equiv W_nR_nW_n^T$.

For the derivation of the Fokker-Planck equation, it is useful to introduce the rotated time delay matrix $Z_n\equiv  J_nQ_nJ_n^T$, for which the recursion relation takes the form
\begin{eqnarray}
&&U_{n}Z_{n+1}U_{n}^T=C_n\left(Z_n+ K_n\right)C_n+ W_nR_nW_n^T,\nonumber\\
&& U_n\equiv W_nJ_n^T,\;U_nU_n^T=I.
\end{eqnarray}
We note that the  matrices $Z_n$ and $Q_n$ have the same eigenvalues. Since $K_n\sim \dell$ and one should keep only the terms up to the first order in $\dell$, we can rewrite the above equation as
\begin{eqnarray}
\label{Z-transform}
U_{n}Z_{n+1}U_{n}^T&=&C_n(Z_n+\Gamma_n R_n\Gamma_n^T)C_n+K_n.
\end{eqnarray}


\section{Fast and slow variables}

It is clear from the structure of Eq.\eqref{S-recursion-suppl} that the main change of the $S$-matrix at each step of the recursion occurs due to the term $e^{i\hq \dell}$, describing the evolution of the $S$-matrix in the absence of disorder. Therefore a typical change of the $S$-matrix is governed by the parameter $k_F\dell=\frac{\dell}{\lfermi}$. At the same time, Eq.\eqref{Q-recursion-suppl} suggests that there are two sources for the change of $\ttau$: the additive one due to the $K$ matrix and the multiplicative one due to the $A^2$ matrix. Using the fact that $\avg{Tr K}=\tr \hq^{-1}\,\dell$ and Eq.\eqref{tras_trqinv} one can see that a typical change of $\ttau$ can be estimated as
\begin{eqnarray}
\delta \ttau \sim \frac{\dell}{\sqrt{E}}+\frac{\dell}{l}\ttau \sim \frac{\dell}{l}(\ttau+\tau_s).
\end{eqnarray}
Since $\frac{\dell}{l}\ll \frac{\dell}{\lfermi}$ for weak disorder, we conclude that $S$ is a fast variable, while $\ttau$ is a slow variable. The eigenvectors of $Q$ are changed primarily due to the $W$ and $J$ matrices, whose definitions involve the $S$-matrix, and thus they also should be considered as fast variables. The compact nature of the fast variables leads to their  randomisation after their total change becomes of the order of unity, implying that the corresponding length scale is of the order of $\lfermi$. 

After such a randomisation of the fast variables occurs their statistics is described by some stationary distribution functions $P_S(S)$ and $P_O(O)$. For the derivation of the Fokker-Planck equation it is sufficient to know only the first two moments of $P_S(S)$ and $P_O(O)$ and additionally the fourth moment of $P_O(O)$. It follows from  Eq.\eqref{S-recursion-suppl} that $P_S(S)$ satisfies the condition
\begin{eqnarray}
P_S(S)=P_S(e^{i\hq \dell}Se^{i\hq \dell})
\end{eqnarray}
and hence
\begin{eqnarray}
\avg{S_{mn}}=\avg{S_{mn}}e^{i(q_m+q_n)\dell},
\end{eqnarray}
implying that 
\begin{eqnarray}\label{Savg}
\avg{S_{mn}}=0.
\end{eqnarray}
In a similar way, 
\begin{eqnarray}
\avg{S_{mn}S_{kl}}&=&\avg{S_{mn}S_{kl}}e^{i(q_m+q_n+q_k+q_l)\dell}, \\
\avg{S_{mn}S_{kl}^*}&=&\avg{S_{mn}S_{kl}}e^{i(q_m+q_n-q_k-q_l)\dell} 
\end{eqnarray}
and therefore
\begin{eqnarray}
\label{SSavg}\avg{S_{mn}S_{kl}}&=&0,\\
\label{SSstaravg}\avg{S_{mn}S_{kl}^*}&=&\frac{1}{N+1}(\delta_{mk}\delta_{nl}+\delta_{ml}\delta_{nk}),
\end{eqnarray}
where the normalisation factor in the last equation was derived from the unitarity of the $S$-matrix and an additional assumption that the variance of all diagonal and all off-diagonal elements is the same. The last assumption is physically equivalent to complete randomisation of the momentum of a reflected particle, which is expected for weak disorder.   

Applying similar arguments to  $P_O(O)$, which must be invariant under the transformation $O\to J_n OJ_n^T$,  we obtain
\begin{eqnarray}
\label{Oavg}\avg{O_{mn}}=0, \quad \avg{O_{mn}O_{kl}}&=&\frac{1}{N}\delta_{mk}\delta_{nl}.
\end{eqnarray}

To derive the Fokker-Planck equation we need to average some expressions containing four matrices $O$. In the leading order in $1/N$ such averages can be calculated applying the Wick's theorem and using Eq.\eqref{Oavg}:
\begin{eqnarray}\label{4O}
\avg{O_{ij}O_{mn}O_{pq}O_{st}}
&=&\frac{1}{N^2}(\delta_{im}\delta_{jn}\delta_{ps}\delta_{qt}+
\delta_{pm}\delta_{qn}\delta_{is}\delta_{jt}\nonumber\\
&&+\delta_{ip}\delta_{jq}\delta_{ms}\delta_{nt})+O\left(N^{-3}\right).
\end{eqnarray}

To support the validity of the above relations we computed $S$ and $Q$ matrices numerically using the recursion relations for the two-dimensional Anderson model. A rectangular lattice with $L=100$ cites in the longitudinal direction and $M=488$ in the transverse direction was considered. The energy $E=0.4$ corresponds to $N=100$ open channels and the disorder is given by the random variable uniformly distributed in the interval $[-W/2,W/2]$  with $W=0.8$. The mean values and the standard deviations were calculated by averaging over $2000$ realisations of the random potential and all matrix elements of the corresponding matrices. The following results for the averages of the matrix elements were obtained:  $|\avg{S_{mn}}|=2.2\times 10^{-4}\pm 1.1\times 10^{-2}$, $|\avg{S_{mn}S_{m(n+1)}}|=3.1\times 10^{-6}\pm 2.2\times 10^{-4}$, $(N+1)\avg{|S_{mn}|^2}=0.99\pm 8.7\times 10^{-2}$, $\avg{O_{mn}}=2.7\times 10^{-3}\pm 2.5\times 10^{-3}$, $|\avg{O_{mn}O_{m(n+1)}}|=4.3\times 10^{-8}\pm 2.2\times 10^{-4}$, $N\avg{O_{mn}^2}=1.0\pm 9.6\times 10^{-2}$,  $|\avg{S_{mn}O_{mn}}|=7.5\times 10^{-7}\pm 2.2\times 10^{-4}$, $\avg{O_{mn}^2O_{m(n+1)}^2}/(\avg{O_{mn}^2}\avg{O_{m(n+1)}^2})=0.98 \pm 4.7\times 10^{-2}$. These results are in agreement with Eqs.\eqref{Savg}, \eqref{SSavg}, \eqref{SSstaravg}, \eqref{Oavg} and \eqref{4O}.


\section{Derivation of the Fokker-Planck equation}

The evolution equation for for the joint probability distribution function $\ptn$ of the eigenvalues $\{\ttau_i\}$ of the matrix $Z_n$ can be derived from the recursion relation for $Z_n$ using the second order perturbation theory.

According to Eq.\eqref{Z-transform}
\begin{eqnarray}
U_{n}Z_{n+1}U_{n}^T&=&Z_n+\dz_n, \quad \dz_n=\dz_n^{(1)}+\dz_n^{(2)}, \nonumber\\
\dz_n^{(1)}&\equiv& C_nZ_nC_n-Z_n+K_n\nonumber \\
\dz_n^{(2)}&\equiv&C_n\Gamma_n R_n\Gamma_n^TC_n.
\end{eqnarray}
One can show that the contribution from $\dz_n^{(2)}$ has an extra factor of $(\lfermi/l)$ compared to the contribution from $\dz_n^{(1)}$ and hence can be neglected for weak disorder. The contribution from $\dz_n\equiv \dz_n^{(1)}$ must be expanded up to the linear term in $\dell$. Recalling that $C_n\equiv (I+\Delta_n)^{-1/2}$ and $\Delta_n$ is defined in Eq.\eqref{Delta-def}, we notice that  $A\sim \vmat$ and $\avg{\vmat}=0$, $\avg{\vmat\vmat'}\sim  \dell$, so that we should keep only the terms up to the second order in $A$. Thus we obtain
\begin{eqnarray}\label{dz}
\dz_n&=&-\frac{1}{2}(\Delta_nZ_n+Z_n\Delta_n)+\frac{1}{4}(\Delta_nZ_n\Delta_n)\nonumber\\
&+&\frac{3}{8}(\Delta_n^2Z_n+Z_n\Delta_n^2)+K_n+O(\dell^2).
\end{eqnarray}
The eigenvalues $\{\ttau_i\}$ of $Z_{n+1}$ are the same as the eigenvalues of $Z_n+\dz_n$.
\begin{eqnarray}
Z_n&=&O_n D_nO_n^T, \quad D_n=\diag (\{\ttau_i^{(n)}\}),\\
Z_n+\dz_n&=&O_n\left(D_n+O_n^T\dz_n O_n\right)O_n^T, \\
\ttaunpo_i&=&\ttaun_i+\delta \ttau_i,
\end{eqnarray}
where $\ttaun_i$ are eigenvalues of $Z_n$ and $O_n$ is the matrix of its eigenvectors.

The joint probability distribution function $\ptnpo$ can be calculated as
\begin{eqnarray}\label{pln-init-suppl}
&&P_{n+1}=\avg{\prod_i\delta(\ttau_i-\ttaunpo_i)}=P_n\nonumber\\
&& -\left[
\sum_i \frac{\partial}{\partial \ttau_i}\avg{\dtau_i}
-\sum_{ik}\frac{\partial^2}{\partial \ttau_i \partial\ttau_k}\frac{\avg{\dtau_i\dtau_k}}{2}
\right]P_n.
\end{eqnarray}
The averages $\avg{\dtau_i}$ and $\avg{\dtau_i\dtau_k}$ can be computed with the help of the perturbation theory.
The first and second order results for the eigenvalues of $D_n$ with the perturbation $O_n^T\dz_n O_n$ are given by
\begin{eqnarray}
\dtau_i=\brai O^T\dz O\keti +\sum_{k\neq i}\frac{|\brak O^T\dz O\keti |^2}{\ttau_i-\ttau_k},
\end{eqnarray}
where $\{\keti\}$ is the standard basis in $\mathbb{R}^N$ and we omit the index $n$ for all variables.

Next the following steps should be taken. One substitutes $Z=O DO^T$ and $\dz$ from Eq.\eqref{dz} into Eq.\eqref{pln-init-suppl} and performs first averaging over the orthogonal matrix $O$ and the scattering matrix $T=e^{2i\Phi}$. For the latter averaging one uses the  relations 
\begin{eqnarray}
\avg{\tr X T Y T^{-1}}_T&=&\frac{1}{N+1}\left(\tr X\ \tr Y+ \tr XY^T\right),\nonumber \\
\avg{\tr XT}_T&=&0,  
\end{eqnarray}
which follow from Eq.\eqref{Savg} and \eqref{SSstaravg} and hold for any matrices $X$ and $Y$. For the averaging over the orthogonal matrix $O$ one has a similar relation 
\begin{eqnarray}
\avg{\tr X O Y O^T}_O=\frac{1}{N}\tr X\ \tr Y,
\end{eqnarray}
following from Eq.\eqref{Oavg}. Expressions involving four matrices $O$ are calculated using Eq.\eqref{4O}.

After the first step we are left only with the terms containing the eigenvalues $\ttau_i$ and two more terms $\tras$ and $\tr\: \hq^{-1}$ coming from the averaging of the terms involving $\Delta$ and $K$ respectively. For the former one we first average over the disordered potential $V(\br)$ and then calculate a sum over transverse momenta $q_{\bn}$. For the latter one we just calculate a sum over $q_{\bn}$. In the leading order in $N$ the results are given by
\begin{eqnarray}\label{tras_trqinv}
\tras&=&\frac{2NA_d \dell}{l},\quad \tr \hq^{-1}=\frac{N A_d }{\sqrt{E}},
\end{eqnarray}
where $A_d\equiv \frac{\sqrt{\pi}\Gamma\left(\frac{d+1}{2}\right)}{ \Gamma\left(\frac{d}{2}\right)}$.

Finally, scaling the variables $\tau=\frac{\ttau}{\tau_s}$ and  $r=A_d\frac{L}{l}$ and taking the limit $\dell\to 0$,
one obtains the Fokker-Planck equation for the distribution function
$P(\{\tau_i\},r)$
\begin{eqnarray}\label{FP-tau-final-const-suppl}
\frac{\partial P}{\partial r}&=&
\frac{1}{N}\sum_i \frac{\partial}{\partial \tau_i}
\left[(N-1)\tau_i-2N\right.\nonumber\\
&&\left.
-\sum_{k\neq i}\frac{\tau_i^2}{\tau_i-\tau_k}+\frac{\partial }{\partial \tau_i}\tau_i^2\right]P.
\end{eqnarray}
Although $1/N$ corrections were discarded in the derivation of this equation, we include such a correction to the first term on the r.h.s. for the following reason.  Studying the moments $\avg{\tau^q}$ directly from the recursion relation \eqref{Q-recursion-suppl}, one can show that the relative values of the $1/N$ corrections to the first and the third terms are constrained. By writing $N-1$ instead of $N$ in the first term one makes sure that this constraint is satisfied. In particular, it guarantees that there are no $1/N$ corrections to the evolution of $\avg{\tau}$, as one expects from Eq.\eqref{Q-recursion-suppl}.    



\begin{thebibliography}{99}
\bibitem{comment} We note that a similar coexistence of a continuous spectrum and an isolated discrete eigenvalue corresponding to the stationary solution was previously found in the $\sigma$-model solution of the quasi-$1d$ problem \cite{FM92}. We also note that Eq.\eqref{u_sol_prelim} can be considered as a generalisation of the one-dimensional result derived in \cite{ACT97}.
\end{thebibliography}
\end{document}